\documentclass[aps,prb,onecolumn,floatfix]{revtex4}

\usepackage{epsf}
\usepackage{subfigure}
\usepackage{amsmath}
\usepackage{amssymb}
\usepackage{bm}
\usepackage{graphicx}
\usepackage{epstopdf}
\newcommand{\mean}[1]{\left \langle #1 \right \rangle}

\newcommand{\be}{\begin{equation}}
\newcommand{\ee}{\end{equation}}
\newcommand{\bea}{\begin{eqnarray}}
\newcommand{\eea}{\end{eqnarray}}

\newcommand{\parent}[1]{\left( #1 \right)}

\newcommand{\tr}{\mathop{\text{tr}}\nolimits}

\begin{document}

\title{\bf Exact expression for the large deviations of mesoscopic currents}

\author{David Andrieux}

\begin{abstract}
Large deviations quantify the occurrence of events that depart from the average behavior of a system. 
In this note we derive an exact expression for their moment generating function. This expression offers a new tool to investigate the behavior of mesoscopic systems.
To illustrate our result, we derive the exact formulae for the current fluctuations in a disordered ring and in a model of ion transport through a membrane.
\end{abstract}

\maketitle

\vskip 0,25 cm

The fluctuations of dynamical and thermodynamical quantities such as energy or matter fluxes are key to understand the behavior of mesoscopic systems, from molecular motors to chemical reaction networks and electronic transport in quantum dots.\cite{H05, S76, EHM09, S12, G13} It is increasingly recognized that fluctuations far away from the average behavior can serve as a probe of the underlying dynamics. \cite{G98, BB00, T09}

The analytical or numerical analysis of large deviations remains, however, challenging. 
Large deviations are governed by a dynamics that does not conserve the probability flows, rendering many traditional techniques unsuitable for their study. 
In addition, large deviations become exponentially rare in time, making their numerical simulation resource-intensive.

We address these difficulties by observing that, while calculating the probability of a current fluctuation is not possible in most cases, the inverse problem of finding the current that corresponds to a given probability is actually tractable. 
This leads to an exact expression for the inverse of its moment generating function, and thus for all the moments of the current fluctuations. 
Our result holds for arbitrary kinetic diagrams, providing a general and efficient mean to explore the transport properties of mesoscopic systems.

\section{Mesoscopic description of fluctuations}

\subsection{Markov chains and thermodynamic description}

We first consider discrete-time Markov chains. Continuous-time Markov processes will be treated in Section \ref{conttime}.

A Markov chain is characterized by a transition matrix $T = \parent{T_{ij}} \in \mathbb{R}^{N\times N}$ on a finite state space.
The operator $T$ is stochastic, i.e., it is non-negative ($T \geq 0$) and its rows sum to one ($\sum_{j} T_{ij} = 1)$.
We assume that the Markov chain is primitive, i.e., there exists an $n_0$ such that $T^{n_0}$ has all positive entries. 
This guarantees that $T$ has a unique stationary distribution $\pmb{\pmb{\pi}}$ such that $\pmb{\pmb{\pi}} = \pmb{\pmb{\pi}} T$.

A Markov chain $P$ defines an equilibrium dynamics when
\begin{eqnarray}
\pi_i T_{ij} = \pi_j T_{ji}  
\label{DB}
\end{eqnarray}
for all transitions. 
These conditions are equivalent to
\bea
T_{i_1 i_2} \ldots T_{i_n i_1} = T_{i_1 i_n} \ldots T_{i_2 i_1 } \quad  \text{for any finite sequence} \; (i_1, i_2, \ldots, i_n) \, .
\label{Kolmogorov}
\eea
Note that the latter conditions do not require the knowledge of the steady state, in contrast to the detailed balance conditions (\ref{DB}).
At equilibrium, no probability flux is present in the stationary state.

Out of equilibrium probability fluxes ciculate through the system and generate thermodynamic forces or affinities. 
The affinities can be measured by the breaking of detailed balance along cyclic paths $c = (i_1, i_2, \ldots, i_n)$ as
\bea
\frac{T_{i_1 i_2} \ldots T_{i_n i_1}}{ T_{i_1 i_n} \ldots T_{i_2 i_1 } } = \exp \parent{A_c} \, .
\label{A}
\eea
Each transition then supports a mean current
\bea
J_{ij} = \pi_i T_{ij} - \pi_j T_{ji}  \, .
\eea
These currents can also be expressed in terms of the currents along the cycles $c$ of the chain.\cite{H05, JQQ04}

%

\subsection{Mesoscopic currents and their fluctuations}

The Markov chain $T$ generates random paths $i_0 \rightarrow i_1 \rightarrow \ldots \rightarrow i_n$.
These paths generate fluctuating currents measured by
\bea
G_l (n) =\sum_{k=1}^n j_l(k) \, ,
\eea
where
$l$ denotes a transition $i \rightleftharpoons j$ and $j_l(k) = \pm 1$ if the transition $i_{k-1}\rightarrow i_{k}$ corresponds to $l$ in the positive (negative) direction, and $0$ otherwise.
 
The large deviations of the current $j_l$ can be characterized by the cumulant generating function
\bea
q_l(\lambda) = \lim_{n\rightarrow\infty} \frac{1}{n} \ln \mean{ {\rm e}^{\lambda G_l(n)}  }
\eea
or, equivalently, by the moment generating function
\bea
m_l(\lambda) = \exp q_l(\lambda) \, .
\label{m.l}
\eea
In particular, all the moments are obtained by successive derivations:
\bea
\mu_k \equiv \frac{ {\rm d}^k m_l (0)}{{\rm d} \lambda^k } = \lim_{n\rightarrow\infty} \frac{1}{n} \mean{G_l^k(n)}.
\label{momentk}
\eea
Similarly, the $k^{{\rm th}}$ derivative of the cumulant generating function $q(\lambda)$ at the origin generates the cumulant of order $k$.

The moment generating function (\ref{m.l}) is given by \cite{AG07}
\bea
m_l(\lambda) = \rho\parent{ T \circ Z_l(\lambda) } \, ,
\label{rho}
\eea
where $\rho (A)$ denotes the spectral radius (i.e. the modulus of the largest eigenvalue) of the operator $A$, $\circ$ is the Hadamar product of two operators, and
\bea
(Z_l)_{ij} (\lambda) \equiv \begin{cases} \exp \parent{+ \lambda/2}  & \text{if the transition $i\rightarrow j$ corresponds to $l$ in the positive direction,} \\
                 \exp \parent{-\lambda/2}  & \text{if the transition $i\rightarrow j$ corresponds to $l$ in the negative direction,} \\
                1 & \text{otherwise}.
                \end{cases}
\nonumber
\eea
The operator $T\circ Z$ is non-stochastic when $\lambda \neq 0$.
Consequently, its spectral radius, and thus the corresponding current fluctuations, cannot be resolved analytically in most cases.
Nonetheless, we proceed to find an analytical expression for the large deviations of the currents.

\section{The new method}

\subsection{Exact expression for the inverse generating function}

Our approach rests on the finding that, while is difficult to obtain $s = m(\lambda)$, calculating the inverse function 
\bea
\lambda (s) = m^{-1} (s)
\label{lambdam}
\eea
is actually tractable (Figure 1).

\begin{figure}[h]
\centerline{\includegraphics[width=12cm]{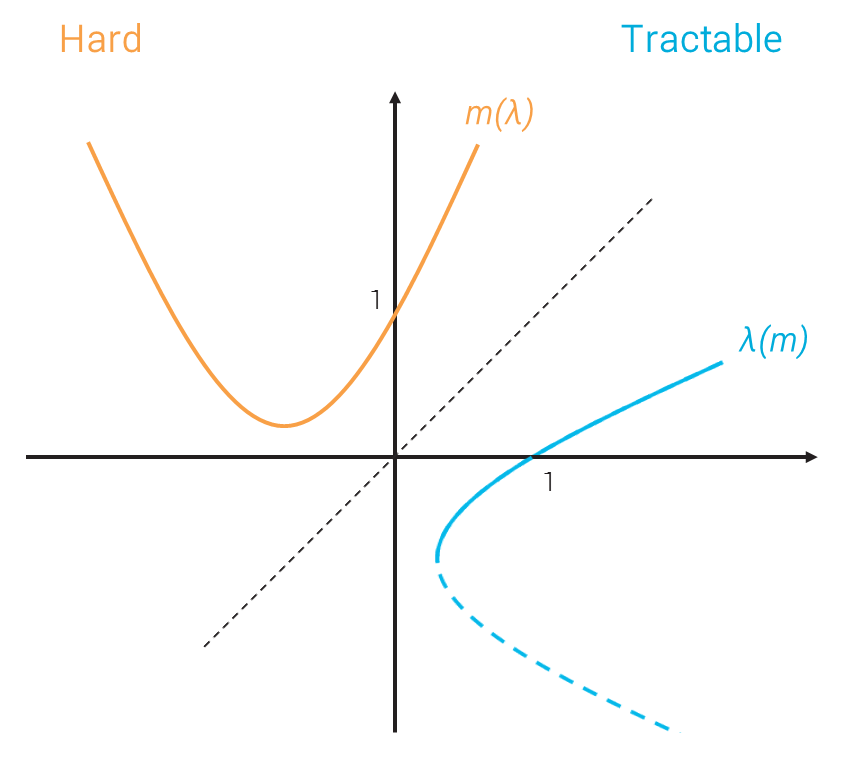}}
\caption{Schematic representation of our method. Calculating the moment generating function (left curve) is not possible in most cases, as it requires finding the leading eigenvalue of a non-stochastic operator.
In contrast, its inverse function (right curve) can be expressed exactly.
}
\label{fig1}
\end{figure}

The generating function $m(\lambda)$ is given by (\ref{rho}) or, equivalently, by the largest solution of the eigenvalue equation 
\bea
\det\parent{sI - Z_l(\lambda)\circ T}=0 \, .
\label{det.eq}
\eea

We will use the cycle decomposition of determinants to analyze this equation. \cite{AG07}
Each term is the determinant corresponds to an element in the group of permutation $S^N$. 
Each such permutation can be decomposed into a product of cycles, each cycle being weighted by the corresponding elements of the operator.
We thus define the weight of a cycle $c = (i_1, i_2, \ldots, i_n)$ as
\bea
w_c = T_{i1 i_2} \ldots T_{i_{n-1} i_n} T_{i_n i_1} \, .
\eea
We also introduce the quantities $D_c(s)$ given by the determinant of $sI-T$ with rows and columns indexed in the complementary set of $c$.

The next step is to isolate the terms containing the counting parameter $\lambda$ in the eigenvalue equation.
To this end we define the passage functions
\bea
\chi_c (i,j) = \begin{cases}1  & \text{if the cycle $c = (i_1 i_2 \ldots i_n)$ contains the transition $i\rightarrow j$ } \, , \\
                0 & \text{otherwise}.
                \end{cases}
\nonumber
\eea
The eigenvalue equation (\ref{det.eq}) can then be decomposed as
\bea
0 = p(s) + \bar{w}_+ (s) \parent{1-{\rm e}^{\lambda}} + \bar{w}_- (s) \parent{1-{\rm e}^{-\lambda}}  \, ,
\label{d.gen}
\eea
where we introduced
\bea
p(s)= \det (sI-T)
\eea
as well as
\bea
\bar{w}_+ (s) = \sum_c \chi_c (i,j) (1-\chi_c(j,i)) \, w_c \, D_c(s)
\label{wp}
\eea
and
\bea
\bar{w}_- (s) = \sum_c \chi_c (j,i) (1-\chi_c(i,j)) \, w_c \, D_c(s)  \, .
\label{wm}
\eea
The factors $\chi  (1-\chi)$ select cycles that contains the transition $l$ in one direction only (i.e, eliminating cycles of the form $(ij)$ that do not contribute to the current).

The factors $\bar{w}_\pm (s)$ can be thought of as 'generalized currents'.
In particular, $w_+ (s) = w_- (s)$ at equilibrium. 
They are positive at $s=1$ so that, by continuity, they remain positive around that point.
Over this range we can write
\bea
0 = p(s) + 2\sqrt{\bar{w}_+ (s) \bar{w}_- (s)} \Big[\cosh\parent{\bar{A}(s)/2} - \cosh \parent{\bar{A}(s)/2 + \lambda}\Big] \, ,
\label{d.gen}
\eea
where the quantities $\bar{w}_\pm$ determine a 'generalized affinity'
\bea
\bar{A} (s) = \ln \frac{\bar{w}_+ (s) }{\bar{w}_- (s) } \, .
\label{A}
\eea
We note that the eigenvalue equation (\ref{d.gen}) presents a fluctuation symmetry $\lambda \rightarrow -\bar{A} - \lambda$ only when the affinity $\bar{A}$ does not depend on $s$.  
The conditions under which individual currents present a fluctuation symmetry can be found in ref. \cite{AG07b}.

Solving eq. (\ref{d.gen}) for the counting parameter $\lambda$ we find
\bea
\lambda_{\pm} (s) = -\frac{\bar{A} (s)}{2} \pm \ln \parent{ x(s)+\sqrt{x^2(s)-1}} 
\label{l.gen}
\eea
with
\bea
x(s) = \cosh\parent{\frac{\bar{A} (s)}{2}} + \frac{p(s)}{2\sqrt{\bar{w}_+ (s)\bar{w}_- (s)}}\, .
\label{x.gen}
\eea
The function $\lambda(s)$ is the inverse of the moment generating function. 

Formulae (\ref{l.gen}-\ref{x.gen}) constitute our central result: they provide an exact expression for the large deviations of general Markov chains.
In the next section we will use this result to derive analytical expressions for the moments of the current fluctuations.

\subsection{Expressions for the moments}

All moments (\ref{momentk}) are obtained by derivating the generating function $m(\lambda)$, whereas we have derived its inverse $\lambda(s)$.
We can relate the derivatives of a function to the derivatives of its inverse by successively differentiating the identity $x = f^{-1}(f(x))$. 
Noting that the point $m(0) =1$ corresponds to $\lambda (1) = 0$, we find 
\bea
\mu_1 = \frac{ {\rm d}m (0)}{{\rm d} \lambda } = \parent{ \frac{ {\rm d} \lambda (1)}{{\rm d} s } }^{-1}= \frac{1}{  \lambda' (1)} \, ,
\label{mu1}
\eea
where the prime denotes a derivative with respect to $s$.

The second moment is in turn given by
\bea
\mu_2 = - \frac{ \lambda''(1)}{\lambda'(1)^3} \, .
\label{mu2}
\eea

We can derive similar relations for higher-order moments.

We now use our result (\ref{l.gen}) to calculate these moments.\footnote{It is actually easier to use the relation (\ref{d.gen}) instead of (\ref{l.gen}) to obtain the moments.}
The point $\lambda (1) = 0$ belongs to the branch $\lambda_+$ when $\bar{A}(1) > 0$ and to the branch $\lambda_-$ when $\bar{A}(1) < 0$. 
However, both cases can be regrouped into a single formulation.
Inserting the derivative of $\lambda_\pm $ into expression (\ref{mu1}), the mean current takes the form
\bea
\mu_1 = \frac{\bar{w}_+ (1) - \bar{w}_- (1)}{p'(1)}\, ,
\eea
where we used that $p(1) = 0$ since the operator $T$ is stochastic.
We can verify that this expression is equivalent to $\pi T_{ij} - \pi_j T_{ij}$ by using the techniques of ref. \cite{JQQ04}.

The next order fluctuations are given by eq. (\ref{mu2}), which becomes
\bea
\mu_2 = \frac{\bar{w}_+(1) + \bar{w}_-(1)}{ p'(1)} -  (\bar{w}_+(1) - \bar{w}_-(1))^2 \frac{p''(1)}{p'(1)^3} + 2  \frac{(\bar{w}'_+(1) -\bar{w}'_-(1) )(\bar{w}_+(1) - \bar{w}_-(1))}{p'(1)^2} \, .
\eea
We see that second-order fluctuations arise from a 'shot noise' contribution that is always present (the first term), and from a second contribution (the last two terms) that only appears out of equilibrium (remember that $\bar{w}_+(s)= \bar{w}_-(s)$ at equilibrium). 

We can now give explicit expressions for the factors $p'$ and $p''$. The function
\bea
p(s) = \det\parent{sI-T} = \sum_{k=0}^N \gamma_k s^k
\label{ps}
\eea
is a polynomial of degree $N$ with coefficients $\gamma_N = 1$, $\gamma_0 = (-1)^N \det(T)$ and
\bea
\gamma_{N-k} = \frac{(-1)^k}{k!}
\left|\begin{array}{ccccc}
   \tr{A} & k-1 & 0 & \dots &   0  \\
   \tr{A^2} & \tr{A} &  k-2 & \cdots &  0   \\
  \vdots & \vdots  & & \ddots & \vdots \\
  \tr{A^{k-1}} & \tr{A^{k-2}} &  & \cdots & 1  \\
  \tr{A^{k}} & \tr{A^{k-1}}  & & \cdots & \tr{A}\\
\end{array} \right| 
 \, . 
\label{coeffp}
\eea
Therefore, we have that 
\bea
\label{pp1}
p'(1) &=& \sum_{k=1}^N k \, \gamma_k \, , \\
p''(1) &=& \sum_{k=2}^N k(k-1) \, \gamma_k \, , \label{ppp1} \\
&...&  \nonumber
\eea
with the coefficients $\gamma_k$ given by (\ref{coeffp}).\footnote{The Faddeev-LeVerrier algorithm computes these coefficients more efficiently.}
We thus have obtained complete expressions for the moments.

\subsection{Extension to continuous time}
\label{conttime}

Our derivation readily translates to the case of continuous-time Markov processes described by a master equation 
\bea
\frac{{\rm d} {\bf p}(t)}{ {\rm d}t} = {\bf p}(t)T \, .
\eea 
In this case the transition matrix obeys $T_{ij} \geq 0$ for $i\neq j$ and $T_{ii}=-\sum_{j\neq i}T_{ij}$. 
Its steady state distribution satisfies $0= {\bf p}_{{\rm st}}T$.

In continuous time the leading eigenvalue of the operator $T\circ Z(\lambda)$ corresponds to the cumulant generating function $q(\lambda)$ rather than to $m(\lambda)$ as in discrete time.
The cycle decomposition of the eigenvalue equation remains identical. 
Consequently, our central result (\ref{l.gen}-\ref{x.gen}) takes the exact same form, except than it now describes the inverse of the {\it cumulant} generating function.

The 'origin point' to obtain the cumulants is centered at $(\lambda = 0, s=0)$.  
The quantities $\bar{w}_\pm(s)$ and the derivatives of $p(s)$ must therefore be evaluated at $s=0$ when calculating the $k^{{\rm th }}$ order cumulants:
\bea
\label{kap1}
\kappa_1 &=& \frac{\bar{w}_+ (0) - \bar{w}_- (0)}{p'(0)}\, ,\\
\label{kap2}
\kappa_2 &=& \frac{\bar{w}_+(0) + \bar{w}_-(0)}{ p'(0)} +  (\bar{w}_+(0) - \bar{w}_-(0))^2 \frac{p''(0)}{p'(0)^3} + 2  \frac{(\bar{w}'_+(0) -\bar{w}'_-(0) )(\bar{w}_+(0) - \bar{w}_-(0))}{p'(0)^2} \, ,
\eea
where $\kappa_1 = \mu_1$ and $\kappa_2 = \mu_2-(\mu_1)^2$ and where we used that $p(0) = 0$ in continuous time. 
The derivatives of $p(s)$ at $s=0$ take the simple form 
\bea
\label{pp0}
p'(0) &=& \gamma_1 \, , \\
p''(0) &=& \gamma_2 \, , 
\label{ppp0} \\
\nonumber
&...& \nonumber
\eea
where the coefficients $\gamma_k$ are given by (\ref{coeffp}).

In summary, all our results hold by remplacing $m(\lambda)$ by $q(\lambda)$, the moments $\mu_k$ by the cumulants of order $k$, and calculating the various quantities at $s=0$.

\section{Examples}

\subsection{Transport on a disordered ring}

Periodic chains model many nonequilibrium systems, from the conductivity of anisotropic organic conductors \cite{ABSO81} to molecular motors \cite{AG08} or enzymetic kinetics \cite{MCB10}.
More generally, cycles constitute the building blocks of more complex systems, both from a dynamical and thermodynamical perspective. \cite{S76} 
The characterization of their dynamics is thus fundamental to understand transport at the mesoscopic scale.

\begin{figure}[h]
\centerline{\includegraphics[width=13cm]{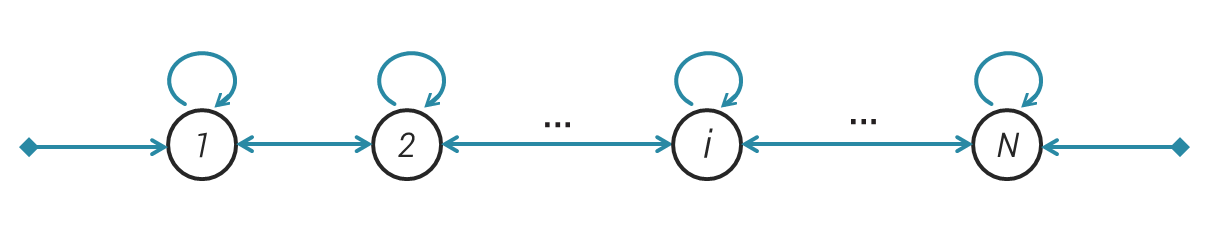}}
\caption{A periodic system of size $N$. The transition probabilities can take arbitrary values.}
\label{fig2}
\end{figure}

We consider a $N$-state system characterized by the allowed transitions $i \rightarrow i$ and $i \rightarrow  i \pm 1$ with periodic boundary conditions. 
The transition probabilities can take arbitrary values (Figure \ref{fig2}).

The current is measured along the transition $i \rightleftharpoons i + 1$. 
Regardless of the choice of $i$, the only non-trivial cycles contributing to the current are $c_+=(1, 2, \ldots, N)$ and $c_-=(N, \ldots, 2, 1)$.
Their corresponding weights are $w_{+} = T_{12}T_{23}\ldots T_{N1}$ and $w_{-} = T_{1N}\ldots T_{32}T_{21}$. 
The affinity $A = \ln w_{+} / w_{-}$ so that the system is at equilibrium when $w_+ = w_-$. 
Because the cycles $c_\pm$ pass through all the states of the system, we have $D(c_+) = D(c_-) =1$

Expressions (\ref{wp}-\ref{A}) therefore simplify to $\bar{w}_\pm (s) = w_\pm$ and $\bar{A}=A$, and are independent of $s$. 
The eigenvalue equation thus reads 
\bea
0 = p(s) + 2\sqrt{w_+ w_-} \Big[\cosh\parent{A/2} - \cosh \parent{A/2 + \lambda}\Big] \, .
\label{dring}
\eea
Note that this equation is invariant under the transformation $\lambda \rightarrow - A - \lambda$. 
This symmetry is at the origin of the fluctuation theorem, according to which $m(\lambda) = m(A+\lambda)$.\cite{AG04, AG07}

\begin{figure}[h!]
\centerline{\includegraphics[width=7cm]{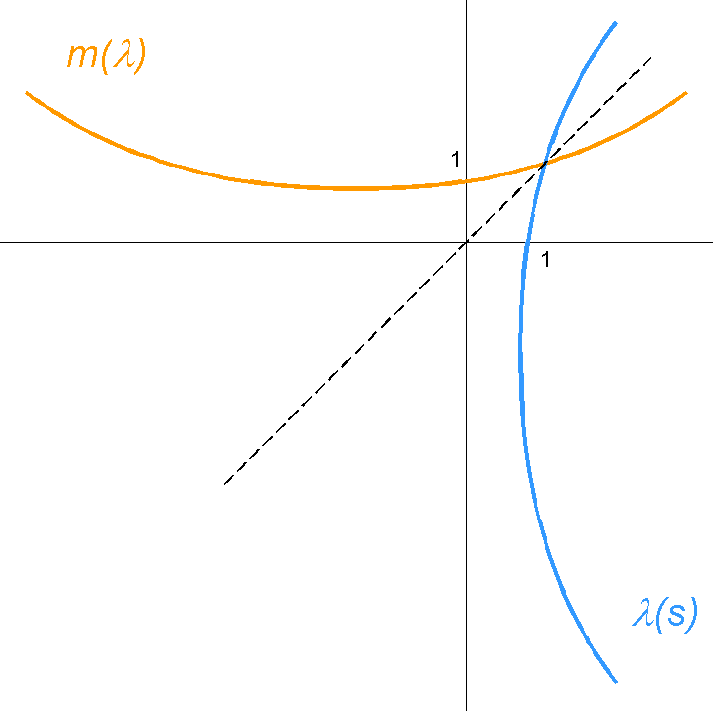}}
\caption{The moment generating function and its inverse for a disordered ring of size $N=3$. 
The moment generating function $m(\lambda)$ is obtained by numerical evaluation of the spectral radius (\ref{rho}) while its inverse function is obtained from the analytical expression (\ref{lcycle}).
The transition probabilities take the values $T_{12} =0.9, T_{13} =0.1, T_{23} = T_{21} = 0.5, T_{31} =0.8, T_{32} =0.2$ and $T_{11}=T_{22}=T_{33} =0$. 
The function $m(\lambda)$ satisfies the fluctuation symmetry $m(\lambda) = m(-\lambda - A)$ with $A = \ln (36)$.
Both functions are mirror images of each other, which reflects their inverse relationship.}
\label{fig4}
\end{figure}

We can solve eq. (\ref{dring}) for the counting parameter $\lambda$. This leads to 
\bea
\lambda_{\pm} (s) = -\frac{A}{2} \pm \ln \parent{ x(s)+\sqrt{x^2(s)-1}}
\label{lcycle}
\eea
with
\bea
x(s) = \cosh\parent{\frac{A}{2}} +  \frac{p(s)}{2\sqrt{w_+ w_-}}\, .
\label{xcycle}
\eea
This formula is illustrated in Figure \ref{fig4}.
This expression remains identical in continuous time, in which case it describes the inverse of the cumulant generating function $q(\lambda)$. 

From expression (\ref{mu1}) the mean current reads
\bea
\mu_1 = \frac{w_+ -w_-}{ p'(1)}\, .
\label{mu1.cycle}
\eea
From expression (\ref{mu2}) the second-order moment takes the form
\bea
\mu_2 = \frac{w_++w_-}{ p'(1)} - (w_+-w_-)^2 \frac{p''(1)}{p'(1)^3} \, .
\label{mu2.cycle}
\eea
The factors $p'(1)$ and $p''(1)$ can be obtained from the formulae (\ref{pp1}) and (\ref{ppp1}).
In continuous time these expressions are evaluated at $s=0$ (eqs. (\ref{pp0}-\ref{ppp0})) and correspond to the cumulants $\kappa_k$ (eqs. (\ref{kap1}-\ref{kap2})).

Higher-order moments can be derived similarly.

\subsection{Transport of ions through a membrane}

\begin{figure}[h]
\centerline{\includegraphics[width=13cm]{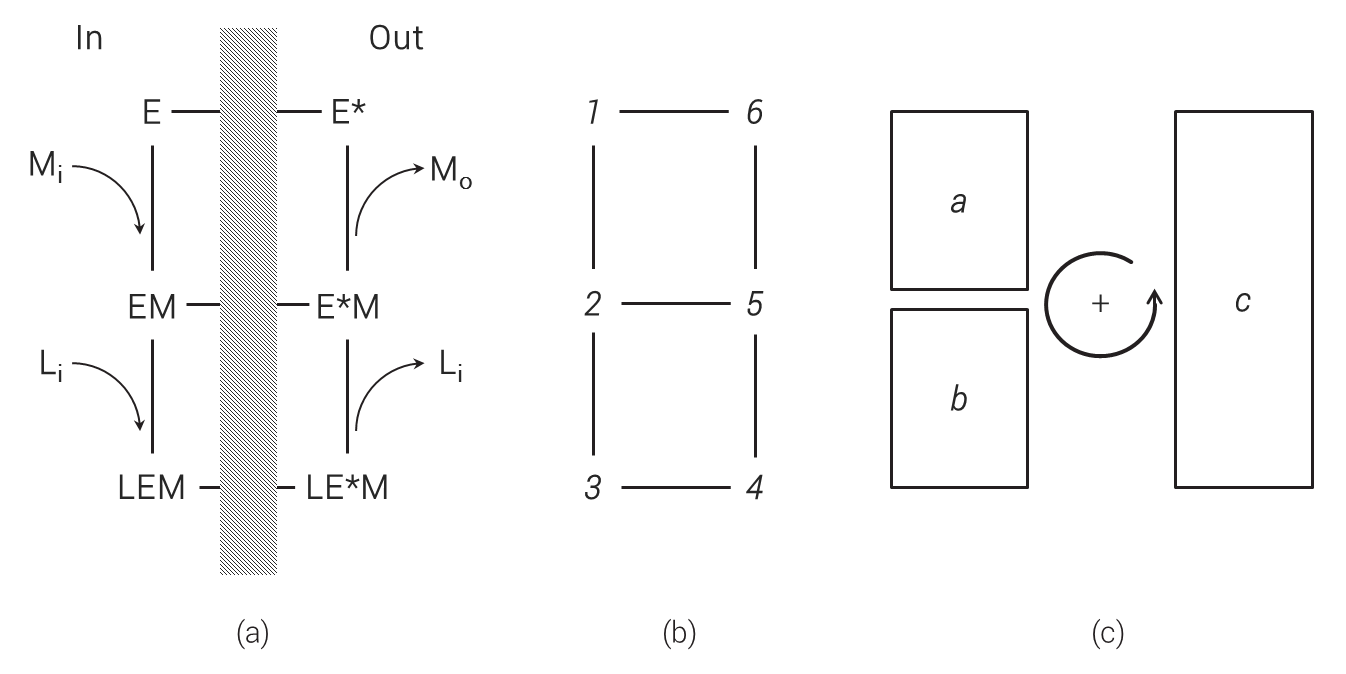}}
\caption{(a) Mechanism for the transport of ions M and L across the membrane. (b) Kinetic diagram. (c) Cycle decomposition. The positive orientation is chosen counterclockwise (adapted from Hill \cite{H05}).}
\label{fig3}
\end{figure}

Following Hill \cite{H05}, we consider a cell surrounded by a membrane that separates the cell's interior (In) from its environment (Out). 
A complex E, which can exist in two conformations E and E*, has binding sites for ions L and M.
These sites are accessible to inside molecules in configuration E only, and to outside molecules in configuration E* only. 
L can be bound only when M is already bound on its site (Figure \ref{fig3}a).

In the normal mode of operation, molecule M has a larger concentration inside than outside, $[{\rm M_i}] > [{\rm M_o}]$, while the opposite holds for molecule L, $[{\rm L_o}] > [{\rm L_i}]$. 
The complex E acts as a free energy transducer and utilizes the M concentration gradient to drive molecules of L from inside to outside against its concentration gradient. 
For example, in the case of the Na/K-ATPase complex M and L would correspond to ${\rm K}^+$ and ${\rm Na}^+$, and transport would be coupled to ATP consumption.

The dynamics is stochastic and described by a probability distribution ${\bf p}(t)$ obeying the master equation
\bea
\frac{{\rm d}{\bf p}(t)}{{\rm d}t} = {\bf p}(t)T
\eea
with transition rates $T_{ij} \geq 0$ and $T_{ii} = -\sum_{j\neq i} T_{ij}$ (Figure \ref{fig3}b).
The transitions $1 \rightarrow 2$, $2\rightarrow 3$, $5 \rightarrow 4$, and $6 \rightarrow 5$ involve the binding of an ion, so that their rates account for the inside and outside ionic concentrations. 
Transitions $2 \rightleftharpoons 5$ and $3 \rightleftharpoons 4$ move an ion from one side of the membrane to the other, so that their rates depend on the difference of potential $V$ across the membrane and on the ionic charges $z_{\rm M}e$ and $z_{\rm L}e$. 
Transport is thus governed by the two affinities $A_{\rm M} = \ln([{\rm M_i}] /[{\rm M_i}]) + z_{\rm M}eV /k_BT$ and $A_{\rm L} = \ln([{\rm L_i}] /[{\rm L_o}]) + z_{\rm L}eV /k_BT$.\cite{H05, AG09}

The currents can be measured by the transition $5 \rightleftharpoons 6$ for ion M and $4 \rightleftharpoons 5$ for ion L.
The transport dynamics can be decomposed in terms of the cycles $a, b$, and $c$ (Figure \ref{fig3}c). 
Cycle $a$ moves one M from inside to outside, cycle $b$ moves one L, and cycle $c$ one M and one L. 
Cycle $c$ thus couples the two ionic currents. 

\begin{figure}[t!]
\centering
\begin{tabular}{cc}
\rotatebox{0}{\scalebox{0.25}{\includegraphics{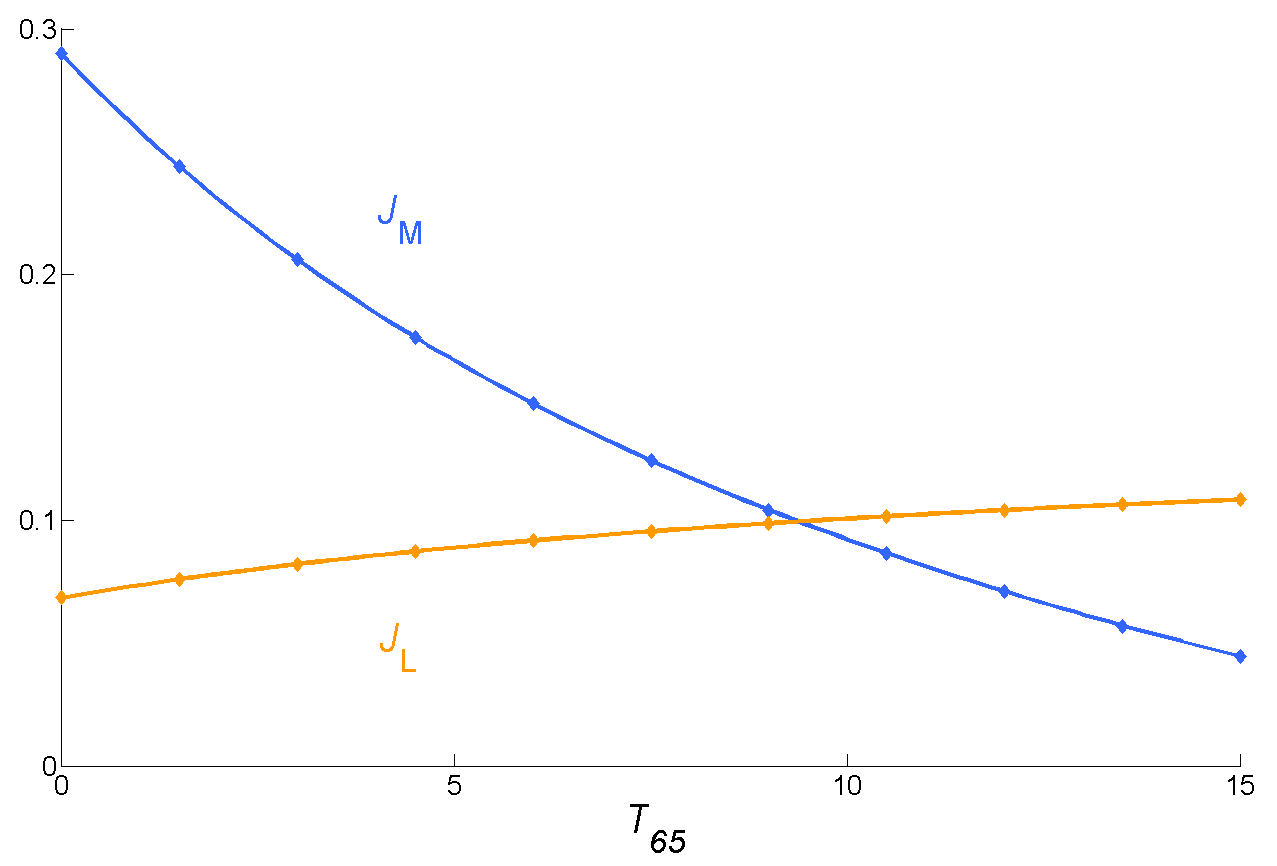}}} &
\rotatebox{0}{\scalebox{0.255}{\includegraphics{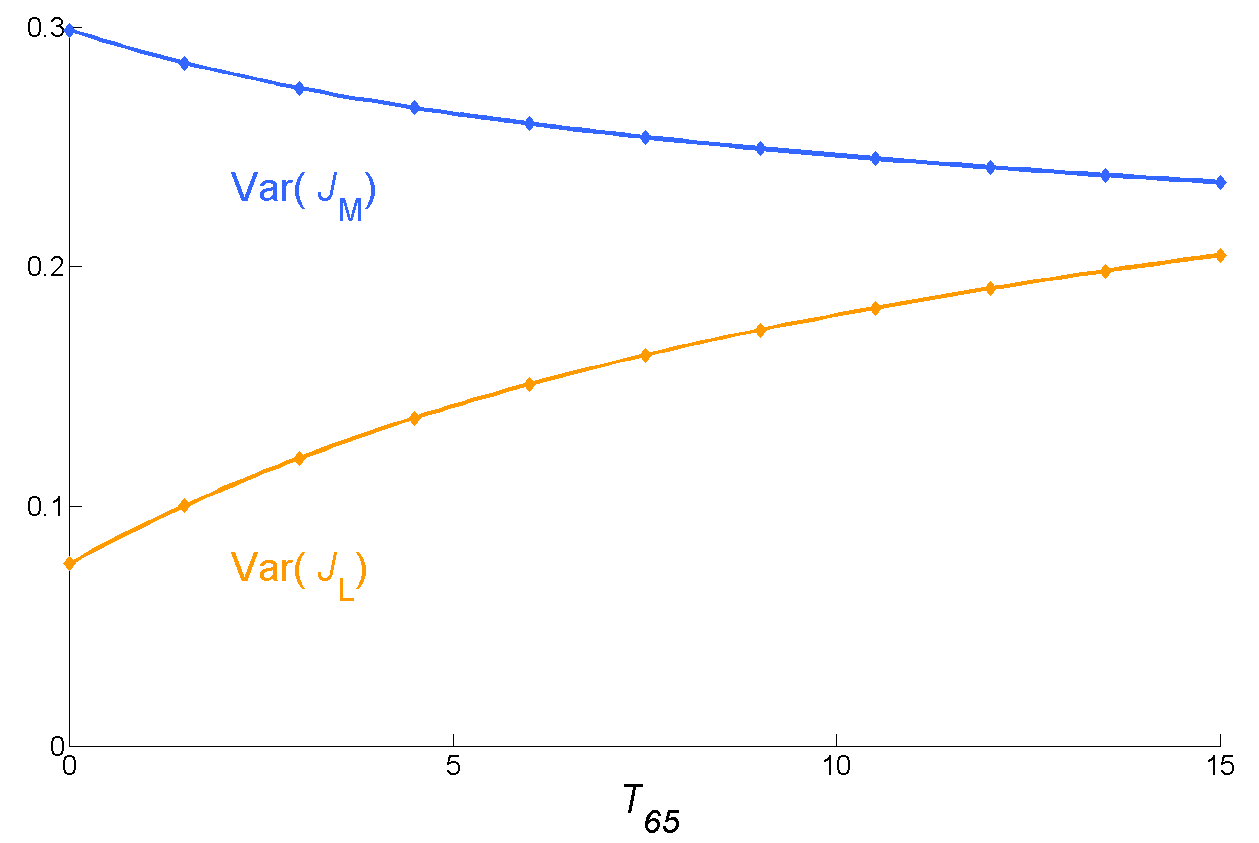}}} \\
\end{tabular}
\caption{Transport properties of the currents of ions M and L as a function of the rate $T_{65}$, which can be adjusted by changing the concentration $[{\rm M_o}]$.
(Left) The mean currents of ions M and L. (Right) The variance of the M and L currents.
The lines depict the formulae (\ref{kap1}) and (\ref{kap2}), which perfectly agree with the results of simulations (diamonds) using Gillespie's algorithm.\cite{G76} 
The transition rates take the values $T_{12} = T_{21} = T_{23} = T_{52} = T_{16}= 2, T_{2,5} = T_{43} = T_{34}= 5, T_{45}=T_{52} = 4, T_{61} = T_{56}=1$, and $T_{ii} = -\sum_{j\neq i} T_{ij}$.
Varying the concentration of one ion affect both currents, here with an opposite effect. Other regimes exist where, e.g., both currents increase or decrease with the concentration $[{\rm M_o}]$.}
\label{fig5}
\end{figure}

 
We first consider the transport statistics of ion ${\rm M}$, whose contributing cycles are $a$ and $c$. 
From (\ref{wp}) and (\ref{wm}) we have that
\bea
\bar{w}_+ (s) = w_{a_+} D_a(s) + w_{c_+}
\eea
and
\bea
\bar{w}_- (s) =w_{a_-} D_a(s) + w_{c_-}  \, ,
\eea
where $w_{c_+} = T_{12}\cdots T_{56}T_{61}$, $w_{a_+} =  T_{12}T_{25}T_{56}T_{61}$, $w_{c_-} = T_{16}T_{65} \cdots T_{21}$, $w_{a_-} =  T_{16}T_{65}T_{52}T_{21}$, and
\bea
D_a (s) = 
\left|\begin{array}{cc}
   s - T_{33} & - T_{34} \\
   - T_{43} & s - T_{44}  \\
\end{array} \right| 
\, . 
\eea
We then readily obtain the mean current and its variance by inserting these expressions into the general formulae (\ref{kap1}) and (\ref{kap2}).
For instance, we have
\bea
J_{{\rm M}} &=& \frac{w_{c_+} - w_{c_-} +(w_{a_+} -  w_{a_-}) D_a(0) }{\gamma_1} 
\eea
with $\gamma_1$ given by (\ref{coeffp}). 
Likewise, we obtain the properties of the current of ion L by replacing $w_a$ by $w_b$ and $D_a$ by $D_b$. 
The resulting mean currents and standard deviations perfectly match our numerical simulations (Figure \ref{fig5}).

The individual currents do not generally obey a fluctuation theorem because their 'effective affinity' $\bar{A} = \ln \bar{w}_+ (s)/\bar{w}_- (s)$  depends on $s$. 
Nonetheless, under certain conditions $\bar{A}$ will be independant of $s$ and a fluctuation theorem for the corresponding current will hold. \cite{AG07b}
For instance, when $w_{a_+}w_{c_-} = w_{c_+}w_{a_-}$, the effective affinity $\bar{A} = \ln w_{c_+}/w_{c_-}= A_{\rm M} + A_{\rm L}$ is independent of $s$ and the current of ion M will obey a fluctuation theorem with this effective affinity. 
The same symmetry holds for the current of ion L when $w_{b_+}w_{c_-} = w_{c_+}w_{b_-}$.\\

This example illustrates how to apply our results to systems in which the transport properties are governed by multiple cycles.
Our results hold both in discrete and continuous time and for arbitrary kinetic diagrams, providing a general and efficient mean to explore the transport properties of mesoscopic systems.



%

\vskip 1.5 cm

{\bf Disclaimer.} This paper is not intended for journal publication. 

%
%
%
%


\end{document}